\documentclass[fleqn,usenatbib]{mnras}

\usepackage[T1]{fontenc}
\DeclareRobustCommand{\VAN}[3]{#2}
\let\VANthebibliography\thebibliography
\def\thebibliography{\DeclareRobustCommand{\VAN}[3]{##3}\VANthebibliography}

\usepackage{graphicx}	% Including figure files
\usepackage{amsmath}	% Advanced maths commands
\usepackage{amssymb}	% Extra maths symbols

\usepackage{newtxtext,newtxmath}
\usepackage{physics,amsmath}

\usepackage[utf8]{inputenc}

\newcommand{\angstrom}{\text{\normalfont\AA}}
\usepackage{newtxtext,newtxmath}

\title[Kinematics of the Local Group gas and galaxies]{Kinematics of the Local Group gas and galaxies in the {\sc Hestia} simulations}

\author[Biaus et al.]{Luis~Biaus$^{1}$\thanks{E-mail: lbiaus@df.uba.ar}, Sebasti\'an~E.~Nuza$^{2, 1}$,
Philipp~Richter$^{3}$,
Martin~Sparre$^{3, 4}$,
Cecilia~Scannapieco$^{1}$,
Mitali~Damle$^{3}$,
\newauthor
Jenny~G.~Sorce$^{4, 5}$,
Robert~J.~J.~Grand$^{6, 7}$,
Elmo~Tempel$^{8, 9}$,
Noam~I.~Libeskind$^{4}$,
Maan~H.~Hani$^{10}$
\\
\\
$^{1}$ Universidad de Buenos Aires, Facultad de Ciencias Exactas y Naturales, Departamento de Física. Buenos Aires, Argentina.\\
$^{2}$ Instituto de Astronom\'{\i}a y F\'{\i}sica del Espacio (IAFE, CONICET-UBA), CC 67, Suc. 28, 1428 Buenos Aires, Argentina\\
$^{3}$ Institut f{\"u}r Physik und Astronomie, Universit{\"a}t Potsdam, Karl-Liebknecht-Str. 24/25, 14476 Golm, Germany\\
$^{4}$ Leibniz-Institut f\"{u}r Astrophysik, An der Sternwarte 16, 14482 Potsdam, Germany\\
$^5$ Universit\'e Paris-Saclay, CNRS, Institut d'Astrophysique Spatiale, 91405, Orsay, France\\
$^6$ Instituto de Astrof\'isica de Canarias, Calle Vía L\'actea s/n, E-38205 La Laguna, Tenerife, Spain\\
$^7$ Departamento de Astrof\'isica, Universidad de La Laguna, Av. del Astrof\'isico Francisco S\'anchez s/n, E-38206, La Laguna, Tenerife, Spain\\
$^8$ Tartu Observatory, University of Tartu, Observatooriumi 1, 61602 T\~oravere, Estonia\\
$^9$ Estonian Academy of Sciences, Kohtu 6, 10130 Tallinn, Estonia\\
$^{10}$ Department of Physics and Astronomy, McMaster University, Hamilton Ontario L8S 4M1, Canada
}

\date{Accepted XXX. Received YYY; in original form ZZZ}

\pubyear{2022}

\begin{document}
\label{firstpage}
\pagerange{\pageref{firstpage}--\pageref{lastpage}}
\maketitle

% Abstract of the paper
\begin{abstract}
We investigate the kinematic properties of gas and galaxies in the Local Group (LG) using high-resolution simulations performed by the {\sc Hestia} (High-resolution Environmental Simulations of The Immediate Area) collaboration. Our simulations include the correct cosmography surrounding LG-like regions %such as the Virgo cluster, the local void and the local filament. The simulated LGs consist
consisting of two main spiral galaxies of $\sim 10^{12}$~M$_\odot$, their satellites and minor isolated galaxies, all sharing the same large-scale motion within a volume of a few Mpc. 
We characterise the gas and galaxy kinematics within the simulated LGs, from the perspective of the Sun, to compare with observed trends from recent HST/COS absorption-line observations and LG galaxy data. To analyse the velocity pattern of LG gas and galaxies seen in the observational data, we build sky maps from the local standard of rest, %and other reference systems such as the 
and the galactic and local group barycentre frames. %, as seen by an observer sitting in the model Milky Way.
%In particular, we focus on the observed velocity dipole seen both in high-velocity absorber samples and galaxy motions towards the barycentre region and its antipodes in the Sky. To analyse the observed trends we build sky maps from the local standard of rest, and other reference systems such as the galactic and local group barycentre frames, as seen by an observer sitting in the model Milky Way. 
Our findings show that the establishment of a radial velocity dipole at low/high latitudes, near the preferred barycentre direction, is a natural outcome of simulation kinematics for material {\it outside} the Milky Way virial radius after removing galaxy rotation when the two main LG galaxies are approaching. Our results favour a scenario where gas and galaxies stream towards the LG barycentre producing a velocity dipole resembling observations. While our study shows in a qualitative way the global matter kinematics in the LG as part of its on-going assembly, quantitative estimates of gas-flow rates and physical conditions of the LG gas have to await a more detailed modeling of the ionization conditions, which will be presented in a follow-up paper.
%However, it is yet to be determined to what extent material in the circumgalactic medium of LG galaxies, which represents roughly half of the gaseous budget in our simulated LGs and is subject to local phenomena such as gas inflows/outflows and/or galactic fountains, could give rise to the observed dipole, at least for the subsample of high-velocity absorbers suggested to reside within the Milky Way halo. 
\end{abstract}


\section{Introduction}

%Our local neighbourhood, the LG\\

The Milky Way (MW) and Andromeda (M31) galaxies are the main constituents of the so-called Local Group (LG) of galaxies, which comprise our immediate cosmological vicinity. In turn, the LG is located within an interwoven network of filaments, sheets and voids, also known as the cosmic web \citep{Bond96,Nuza10,Nuza14b}.
%Together with several dozen of minor galaxy members, these two gian spirals, the Milky Way (MW) and M31, constitute the so-called Local Group (LG) of galaxies, which comprise our immediate vicinity. 
%At larger scales, the LG is located within an interwoven network of filaments, sheets and voids, also known as the cosmic web \citep{Bond96,Nuza10,Nuza14b}.
%LG formation scenario
In the currently accepted cosmological scenario, structures grow in a hierarchical process, with smaller systems merging to form more massive ones. At the same time, galaxies are expected to evolve via accretion and/or condensation of gas onto the potential wells of matter aggregations that previously decoupled from the cosmological expansion \citep[e.g.,][]{White78,Keres05}.

\begin{figure*}
	\includegraphics[width=1.65\columnwidth]{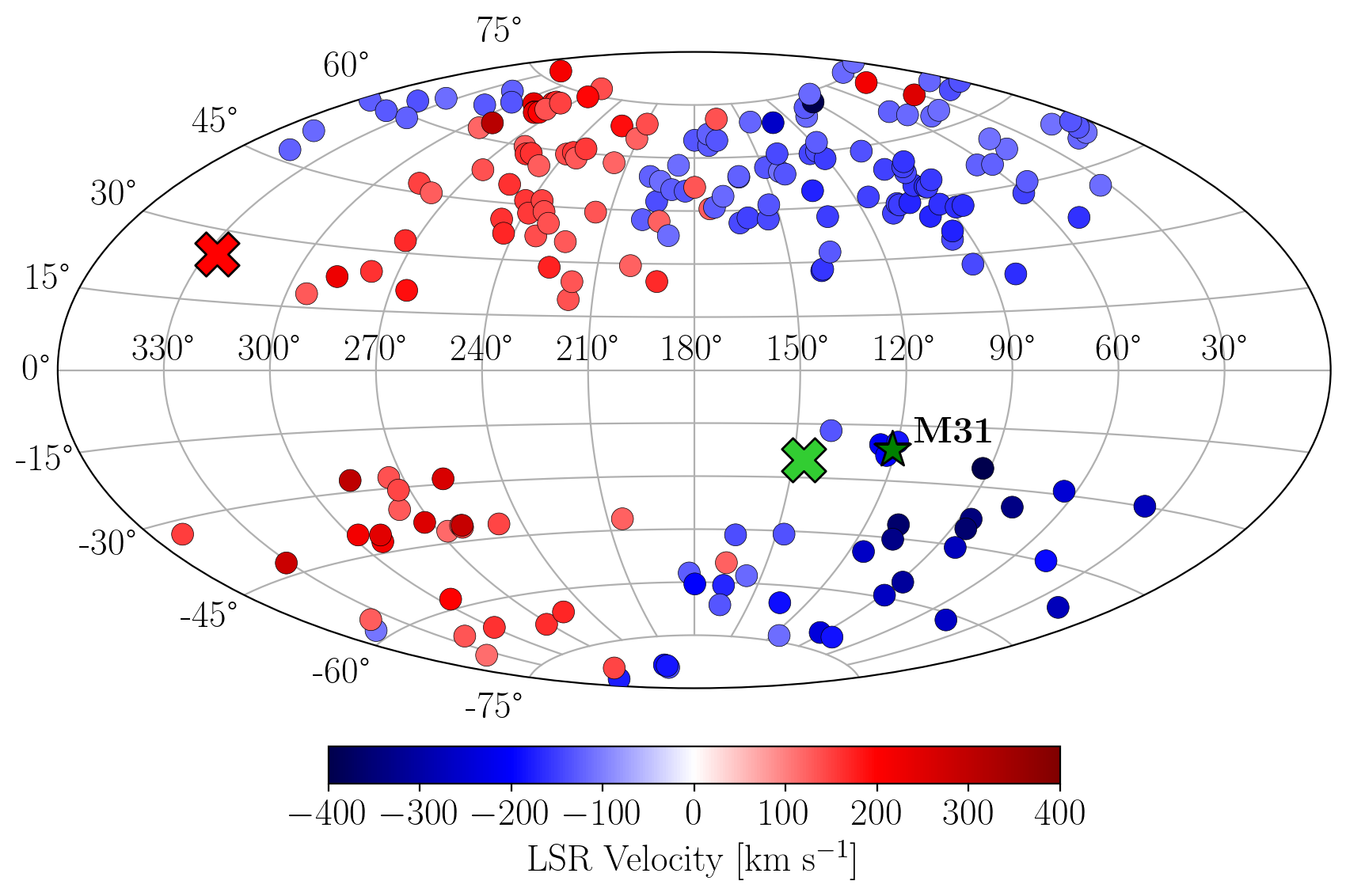}
    \caption{Sky distribution of COS sightlines in a Hammer-Aitoff projection of Galactic coordinates. The green star marks the angular position of M31 while the green (red) cross indicates an observational estimate of the barycentre (anti-barycentre) position, as shown in R17 and references therein. The average LSR gas velocity in each sightline is colour-coded. A velocity dipole is observed even at higher latitudes, where Galactic rotation does not play an important role.
    }
    \label{fig:COS_lines}
\end{figure*}

The cosmic web harbours a major reservoir of baryons that can be accessed through absorption-line measurements in the spectra of background sources \citep[][]{Nicastro18}, Sunyaev-Zel'dovich observations of filaments \citep[][]{deGraaff19}, X-ray emission near galaxy cluster filamentary structures \citep[][]{Eckert15} and, recently, electron dispersion measures of localized fast radio bursts \citep[][]{Macquart20}. At LG scales, ultraviolet (UV) and X-ray absorbers revealed by the spectra of distant galaxies have pointed out the existence of giant multiphase gas haloes surrounding MW and M31 galaxies \citep[e.g,][]{Gupta12,Lehner15,Richter17,Tumlinson17} and, possibly, of LG gas located outside the virial radius of the MW \citep[][]{Wakker03,Sembach03,Collins05,Richter17,Bouma19}. The presence of gas within the LG will, therefore, affect galaxy properties through the astrophysical processes circulating gaseous material in and around galaxies, e.g. in the circumgalactic medium (CGM) and its adjacent regions \citep[][]{Nuza14a,Tumlinson17,Damle22}.  

It is well known that our Galaxy and M31 are on a collision course expected to occur within the next few Gyr, ultimately giving rise to the formation of an elliptical galaxy \citep[e.g,][]{BT08,Cox08,vanderMarel12,Salomon21}. 
%This scenario has been recently confirmed by stellar photometry measurements of the proper motion of M31 \citep{Salomon21}
Actually, the approaching relative velocity results from the general motion of LG galaxies towards the group's barycentre, with MW and M31 being part of the flow \citep[][]{Peebles89,Peebles01,Peebles11,Whiting14}. Since two approaching galaxies drag their own CGM through the intragroup medium (IGrM), it is then natural to assume that galaxy motions and interactions will have an impact on the distribution and kinematics of the surrounding gas. In particular, if the galaxies happen to be in a fairly radial orbit, and/or they are significantly close to each other, the latter will be most noticeable along the direction joining the two systems \citep[see e.g.,][]{Nuza14a,Sparre22}. In the LG, this preferred direction corresponds to that of the LG barycentre, which lies, in projection, close to M31's angular position in the sky, owing to the fact that MW and M31 are the dominant members of the LG \citep[e.g.,][]{Whiting14}.

\cite{Richter17} (hereafter R17) studied a large sample of high-velocity absorbers drawn from archival UV spectra of extragalactic background sources, arising from diffuse gas in the CGM of the MW and in the LG. These observations were obtained with the Cosmic Origins Spectrograph (COS) onboard the Hubble Space Telescope (HST) using a set of gratings in the wavelength range $\lambda = 1150-1775\,\angstrom$ at a spectral resolution of $R \approx 15000-20000$, thus covering the most relevant transition lines of low, intermediate and high ions of silicon and carbon. The sky distribution of all absorbers in R17 is shown in Fig. \ref{fig:COS_lines}, colour-coded by the mean local standard of rest (LSR) velocity of the observed transitions along each sightline. Most of these high-velocity absorbers are linked to known H\,{\sc i} gas complexes such as the Magellanic Stream, Complex A, Complex C and others \citep{Wakker01, Wakker03, Wakker07, Wakker08, Thom06, Thom08, Richter06, Richter16, Richter17}. However, some of the absorbing material towards the {\it general} barycentre direction and its antipode in the sky shows very little H\,{\sc i} $21\,$cm line emission, forming, respectively, a high-velocity dipole at low and high galactic latitudes that they interpret as gas streaming towards the LG barycentre \citep[see also][]{Sembach03,Wakker03,Collins05}. 
Moreover, owing to the low/high latitudes of the absorbers, the observed velocity dipole cannot be explained by Galactic rotation alone. Based on this work, \cite{Bouma19} selected a subsample of absorbers towards the barycentre/anti-barycentre regions and estimated the physical conditions and possible location of the absorbing material by modelling the gas ionization conditions using the column densities of low and intermediate ions as an input. They conclude that a non-negligible fraction of the negative (i.e. approaching) high-velocity clouds lying close to the LG barycentre are most likely gas filling the IGrM outside the virial radius of the MW. This finding opens up the possibility that at least part of the observed dipole at high latitudes is linked to absorbing gas in the IGrM. However, it is necessary to remark that separating absorbers within and beyond MW's CGM is a non-trivial task because they overlap both in velocity and position on the sky. Therefore, the possibility that the observed dipole is actually the result of foreground material within the Galactic CGM cannot be ruled out.

%where it can be seen that the Sun's rotation around the MW imprints a characteristic dipole pattern such that absorbers with $l \in (0^{\circ}, 180^{\circ})$ get blue-shifted and absorbers with $l \in (180^{\circ}, 360^{\circ})$ get red-shifted. 
In relation to this, R17 proposed a scenario where galaxies and gaseous material (connected or unconnected to the galaxies) are falling towards the LG barycentre. In this simple picture, LG gas and galaxies that lie close to the anti-barycentre direction are expected to lag behind the MW flow speed, thus having positive radial velocities, while gas and galaxies located on the opposite side of the LG barycentre will move towards the MW, thus having negative radial velocities. This scenario would naturally explain the velocity dipole observed at high latitudes towards the general barycentre/anti-barycentre direction if some of the absorbers were, in fact, located outside the MW's CGM. Interestingly, this interpretation is consistent with the gas kinematics predicted by the LG constrained cosmological simulation of  \cite{Nuza14a}. 
%(see \cite{Richter17}, their Section 7.3). 
 
During the last decade, a significant effort has been made to simulate the formation of the LG using constrained cosmological simulations within the {\sc Clues}\footnote{\url{www.clues-project.org}} (Constrained Local UniversE Simulations) collaboration by implementing a two-fold program: (i) reproducing the local cosmography after constraining the distribution of matter in the local Universe at the largest scales \citep{Libeskind10, Gottloeber10, Yepes14, Sorce18}, and (ii) selecting LGs from the constrained simulated volumes that fulfill a set of desirable criteria consistent with current observations at Mpc scales \citep[e.g.,][]{Doumler13-I,Doumler13-II, Doumler13-III,Nuza14a,Scannapieco15,Creasey15,Carlesi16,Carlesi20,Libeskind20,Damle22}. As a result, the correct cosmological environment surrounding the simulated LGs accordingly affects their mass assembly history \citep[][]{Carlesi20}. Moreover, after being selected on the basis of having two interacting MW-mass galactic haloes as the main group members, these simulations are able to capture the most salient features of LG-like systems.

In this work, we want to expand the previous work of R17 and study the predicted kinematics of gas and galaxies seen from the Sun using the new generation of LG simulations belonging to the {\sc Hestia} (High-resolution Environmental Simulations of The Immediate Area) project \citep{Libeskind20}. This new simulation set provides high-resolution re-simulations of the LG that improves on several aspects in comparison to previous generations. On the one hand, the initial conditions in {\sc Hestia} are more accurate at reproducing the observed local structure surrounding the LG at the present time. On the other, the simulations are performed using the state-of-the-art, moving-mesh {\sc Arepo} code \citep[][]{Weinberger20} and adopt the {\sc Auriga} galaxy formation model \citep[][]{Grand17}. So far, these simulations have been used to study several aspects of galaxy formation and evolution in the LG such as the modelling of MW and M31 CGM \citep{Damle22}. The major goal of this paper is to characterize the typical flow-pattern of the IGrM in simulated LG analogues from the perspective of the Sun and to evaluate the plausibility of the R17 dipole scenario as described above.

The paper is organized as follows. In Section~\ref{sec:simulations}, we briefly present the main aspects of the LG simulations analysed here. In Section~\ref{sec:analysis}, we describe the setup for the velocity reference frames implemented at Sun's position in the simulated MWs. We also show the spherically-averaged gas mass profiles within the LG, and beyond, including/excluding virialized material. In Section~\ref{sec:results}, we present the predictions for the kinematics of gas and galaxies in the simulations. Finally, in Section~\ref{sec:discussion}, 
we summarise this work and discuss our results in the light of current observations.

\section{Simulations}
\label{sec:simulations}

The {\sc Hestia} simulations are a set of runs aiming at obtaining galaxy systems resembling the LG within a cosmological context. In the following section we present a summary of their main characteristics. For further details we refer the reader to \cite{Libeskind20} and references therein.

\subsection{The simulation code}

The simulations are performed using the moving-mesh {\sc Arepo} code \citep[][]{Springel10,Weinberger20} which computes the joint evolution of gas, stars and dark matter (DM) by solving the gravitational and ideal magnetohydrodynamics (MHD) equations coupled to the {\sc Auriga} galaxy formation model \citep[][]{Pakmor13, Grand17}.
%Magnetohydrodynamics are treated using the moving-mesh code {\sc Arepo} and the {\sc Auriga} galaxy formation model is implemented.
In {\sc Arepo}, a Voronoi mesh is generated from a set of mesh-generating points, which move with the gas velocity. This dynamic mesh reduces the mass flux over interfaces, minimizing advection errors compared to static mesh codes. The MHD equations on the mesh are coupled to self-gravity and other source terms via operator splitting. Magnetic fields are seeded uniformly at $z = 127$ with a comoving field strength of $10^{-14}\,$G.

\begin{table}
\centering
\begin{tabular}{||c c c c c c c||} 
 \hline\hline
 {Simulation} & \multicolumn{2}{c}{$37\_11$} & \multicolumn{2}{c}{$9\_18$} & \multicolumn{2}{c}{$17\_11$} \\ 
 {Galaxy} & MW & M31 & MW & M31 & MW & M31\\
 %\midrulehttps://www.overleaf.com/project/60ba7e76aa0eeb3143ceab3f
 \hline
 $M_{200}$ ($10^{12}\,$M$_{\odot}$) & 0.99 & 1.00 & 1.88 & 2.06 & 1.89 & 2.23\\ 
 $M_{\rm gas}$ ($10^{10}\,$M$_{\odot}$) & 5.76 & 7.71 & 15.1 & 14.7 & 10.5 & 16.3\\
 $M_{\star}$ ($10^{10}\,$M$_{\odot}$) & 5.79 & 5.43 & 10.8 & 12.8 & 11.4 & 12.6\\
 $R_{200}$ (kpc) & 205 & 206 & 254 & 262 & 255 & 269\\ 
 $v_{\rm rad}$ (km s$^{-1}$) & \multicolumn{2}{c}{$9$} & \multicolumn{2}{c}{$-74$} & \multicolumn{2}{c}{$-102$}\\
 $d$ (kpc) & \multicolumn{2}{c}{$850$} & \multicolumn{2}{c}{$866$} & \multicolumn{2}{c}{$675$}\\ [1ex] 
 \hline\hline
\end{tabular}
\caption{Properties of the MW and M31 candidates in the {\sc Hestia} high-resolution simulations. $R_{200}$ is the distance where the mass density profile of a given halo equals 200 times the critical density of the Universe. All masses are measured within this region.}
\label{table:1}
\end{table}

The {\sc Auriga} galaxy formation model implements the main physical processes responsible for the formation and evolution of galaxies within a cosmological context, namely: primordial and metal-dependent gas cooling, a redshift-dependent UV background \citep[][]{FaucherGiguere09}, star formation using a Chabrier initial mass function \citep[][]{SpringelHernquist03,Chabrier03} and chemical/energy feedback from core-collapse/Type Ia supernovae, asymptotic giant branch stars and active galactic nuclei. Galactic winds are created by isotropically launching particles to the local medium from the sites of star formation. The model also follows the formation and evolution of supermassive black holes. For further details, we refer the reader to the works of \cite{Grand17} and \cite{Weinberger20}.

\subsection{The simulated Local Groups}
\label{sec:sim_LG}

\begin{figure*}
	\includegraphics[width=1.95\columnwidth]{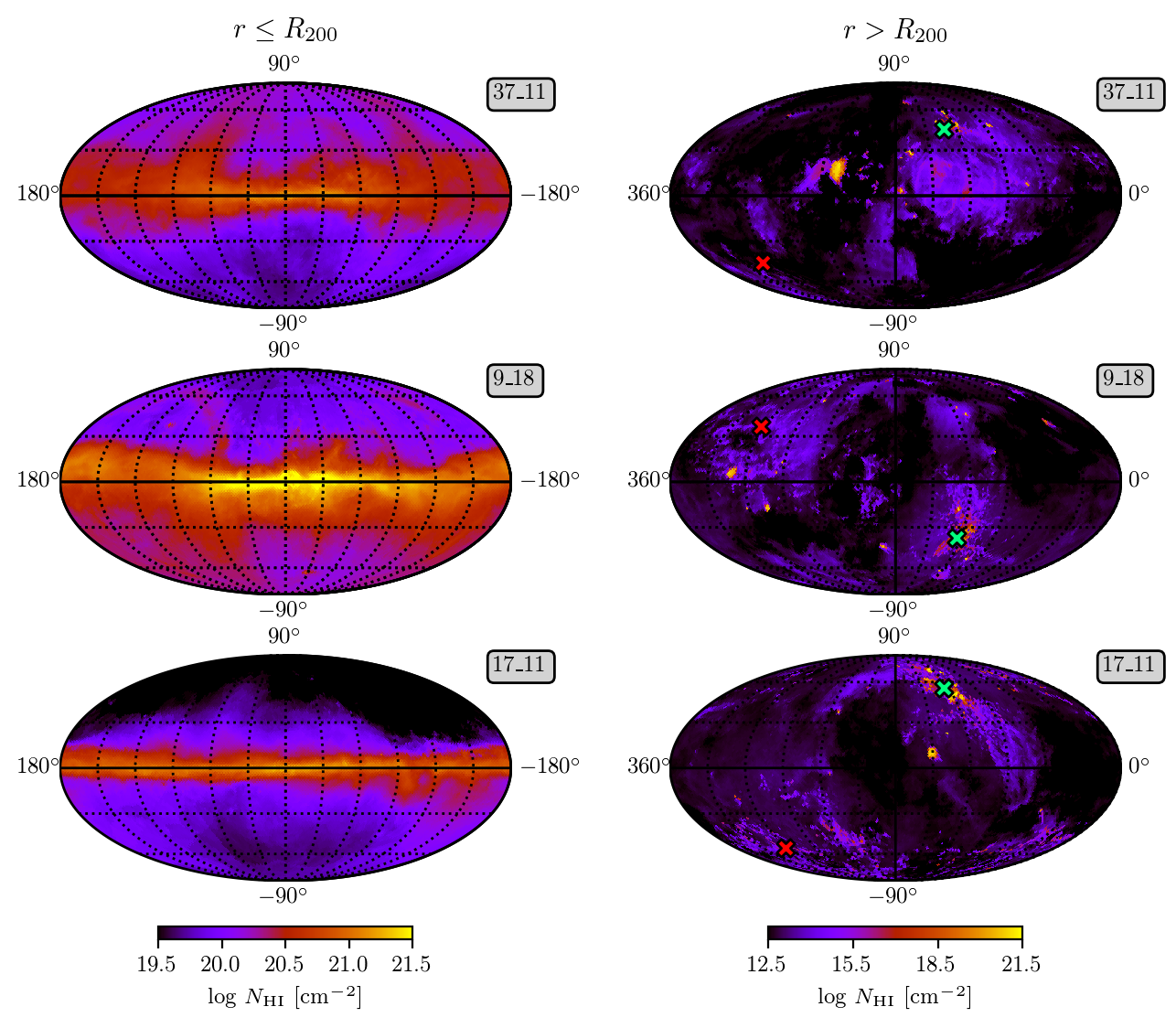}
    \caption{Mollweide projection in Galactic coordinates of H\,{\sc i} column density {\it inside} (left panel) and {\it outside} up to $1000$~kpc; right panel) the virial radius of the MW for the three {\sc Hestia} high-resolution runs as seen from the LSR: $37\_11$ (upper row), $9\_18$ (middle row) and $17\_11$ (lower row). Simulated M31 galaxies are located in the first quadrant of the sky maps for realizations $37\_11$ and $17\_11$, and in the third quadrant for $9\_18$. The very bright spot in the second quadrant of $37\_11$ corresponds to a close satellite. In our simulations, M31 galactic latitudes are determined by the disc plane orientation of each MW counterpart. Crosses indicate the barycentre (green) and anti-barycentre (red) directions in each simulation. 
    }
    \label{fig:nh_density}
\end{figure*}

The simulations used in this work assume cosmological parameters consistent with the best fit of \cite{Planck14}: $\Omega_{\Lambda}=0.682$, $\Omega_{\rm M}=0.270$, $\Omega_{\rm b}=0.048$, $\sigma_8=0.83$ and $H_0=100\,h$\,km\,s$^{-1}$\,Mpc$^{-1}$ with $h=0.677$.

To construct the initial conditions (ICs), observational data from galaxies in our cosmic neighbourhood is taken from the CosmicFlows-2 catalogue \citep{Tully13} to recreate the local Universe's large-scale structure and peculiar velocities. The latter consists of around 8000 peculiar velocities from galaxies with direct-distance measurements up to about $250\,$Mpc from the MW and a median distance of $\sim 70\,$Mpc. From these constraints, about a thousand of DM-only, low-resolution ICs are generated within periodic boxes of $100\,$Mpc\,$h^{-1}$ on a side filled with $256^3$ DM particles. Using a zoom-in technique, a central region encompassing a sphere of radius $10$\,Mpc\,$h^{-1}$ is populated with $512^3$ effective particles of mass $m_{\rm DM} = 6 \times 10^8\,$M$_{\odot}$, replacing their counterparts in the low-resolution run. This is enough to search for LG candidates, as there will be a few thousand particles conforming each of the two main LG haloes. 

The main cosmographic features that the simulations aim to reproduce that are required by the algorithm are: the Virgo cluster, the local void and the local filament. Aside from requiring a Virgo cluster of mass greater than $2 \times 10^{14}\,$M$_{\odot}$ within a distance of $7.5\,$Mpc of an expected region (resulting in a distance of $15-20$~Mpc from the LG pair), the algorithm searches for LG-like regions resembling the actual MW-M31 system that contain halo pairs with mass ratios smaller than 2. As a general rule, the more massive halo is termed `M31' and the other one `MW' in all realizations. %with the following properties: 
\iffalse
\begin{enumerate}
    \item halo masses: $8 \times 10^{11} < M_{\rm halo}/$M$_{\odot} < 3 \times 10^{12}$
    \item separation: $0.5 < d_{\rm sep}/{\rm Mpc} < 1.2$
    \item isolation: no third halo more massive than the smaller one within $2\,$Mpc of the midpoint
    \item halo mass ratio of smaller to bigger halo: $> 0.5$
    \item infalling: $v_{\rm rad} < 0$
\end{enumerate}
\fi
Once the potential LG-like regions have been selected, the ICs are regenerated at a higher resolution using $4096^3$ effective particles within a sphere of radius $5\,$Mpc\,$h^{-1}$, which are subsequently split into a DM-gas cell pair that respects the cosmic baryon fraction. %The mass and spatial resolution of the simulations at this zoom level are $m_{\rm DM} = 1.2 \times 10^6\,$M$_{\odot}$, $m_{\rm gas} = 1.8 \times 10^5\,$M$_{\odot}$ and $\epsilon = 340\,$pc.
From this simulation sample, only three ICs were rerun at an even higher resolution, comprising the ones analysed in this work. These are named $9\_18$, $17\_11$ and $37\_11$ after their corresponding seeds for the long and short waves in the matter density field. In these, two overlapping spheres of diameter $2.5\,$Mpc\,$h^{-1}$ are drawn around the two main LG members and populated with $8192^3$ effective particles. The resulting mass and spatial resolution achieved by these high-resolution runs are $m_{\rm DM} = 1.5 \times 10^5$~M$_{\odot}$, $m_{\rm gas} = 2.2 \times 10^4$~M$_{\odot}$ and $\epsilon = 220$~pc. Throughout this work, we will often sort the simulations by ascending infall velocity between the two main haloes, i.e.: $+9, -74, -102\,$km\,s$^{-1}$ for realizations $37\_11$, $9\_18$ and $17\_11$, respectively\footnote{Note that, in $37\_11$, the two main LG haloes are receding from each other at $z=0$. We stress, however, that the haloes are close to turnaround.}. In Table~\ref{table:1} we summarize the main properties of the MW and M31 candidate galaxies for each of the three high-resolution {\sc Hestia} simulations. Further details concerning the simulated LGs can be found in \cite{Libeskind20}.  
A final remark is that the simulated LGs do not resemble in full detail the observed CGM of the MW. In particular, there is no Magellanic System analogue at a distance of about $50\,$kpc in this set of high-resolution simulations. The Magellanic Stream is a major complex that contributes to a significant portion of the absorbers portrayed in Fig. \ref{fig:COS_lines}, especially towards the direction of M31. Instead, galaxy satellites in the simulations with a comparable mass to the Large Magellanic Cloud that might produce stream-like features lie at larger distances, although still within $R_{200}$ of the simulated MWs. With this caveat in mind, we proceed to study the velocity pattern of gas flows in the simulated LGs which, in the present work, will mainly focus on the kinematics of material outside the simulated CGMs. In a follow-up paper we will come back to this issue.

%\section{Multiphase description}

%Describir las distintas fases y el cálculo de las densidades columnares

\section{Analysis}
\label{sec:analysis}

\subsection{Placing the observer}
\label{sec:LSR}

To determine the relative position of the Sun with respect to the centre of the M31 galaxy candidate in the simulations we need to define an appropriate Galactic coordinate system.
Therefore, we compute the angular momentum of the MW stellar disc and take the north galactic pole in the opposite direction, as required by the usual definition of Galactic coordinates\footnote{As a reminder, we note that the galactic longitude $l$ increases counter-clockwise as viewed from the north galactic pole with $l=0^{\circ}$ corresponding to the direction of the galactic centre.}. The observer is then placed in the midplane of the simulated MW (which defines the galactic {\it latitude} $b=0^{\circ}$) at a distance of $8\,$kpc from its centre and at an azimuthal angle %corresponding to the galactic {\it longitude} $l=0^{\circ}$ 
from which the longitude of the simulated M31 matches the observed one. It is worth mentioning that the {\it actual} latitude of Andromeda is not accurately reproduced in the simulations as the disc plane orientation of the galaxies is inherently random. However, we do not expect here a perfect match to the actual LG but an overall agreement of its main characteristics and large-scale environment. Our goal is to study the kinematics of gas and galaxies in these simulations aiming at performing a comparison of the simulated trends to observations.

% Projection of Sun's velocity vector in the direction of M31:

%IDL> print,sin(121.*!PI/180.)*cos(-21.*!PI/180.)
%     0.800235 (actual M31's position)
%IDL> print,sin(121.*!PI/180.)*cos(-35.*!PI/180.)
%     0.702150
%IDL> print,sin(121.*!PI/180.)*cos(-50.*!PI/180.)
%     0.550977

%Try sitting in the other galaxy to correct for wrong latitude sign of "M31" when necessary? 

Finally, when considering the motion of the Sun around the MW centre in the simulations, the position of the simulated observer defines our LSR reference system. For the Sun's velocity vector we take the galaxy's circular velocity at the observer's radius ($|\vectorbold*{v}_{\odot}| = 220$, $247$ and $218\,$km\,s$^{-1}$ for realizations $37\_11$, $9\_18$ and $17\_11$, respectively\footnote{See Fig.~8 of \cite{Libeskind20}.}) pointing towards $(l,b)=(90^{\circ},0^{\circ})$. For many applications, assuming that the Sun's velocity around the galaxy is purely tangential is a very good approximation. We will also refer to the galactic standard of rest (GSR) reference system which is the same as the LSR but excluding the velocity field produced by the rotation of the galaxy. The LG standard of rest (LGSR), additionally excludes the radial motion of the MW with respect to the LG barycentre \citep[e.g.][]{Karachentsev96,Courteau99}. Specifically, the equations used to derive velocities in each of the mentioned frames from the simulations are:

\begin{equation*}
\vectorbold*{v}_{\rm LSR} \equiv \vectorbold*{v}_{\rm g} - \vectorbold*{v}_{\rm MW} - \vectorbold*{v}_{\odot} + \vectorbold*{v}_{\rm H},
\end{equation*}

\begin{figure*}
    \centering
    \includegraphics[width=2\columnwidth]{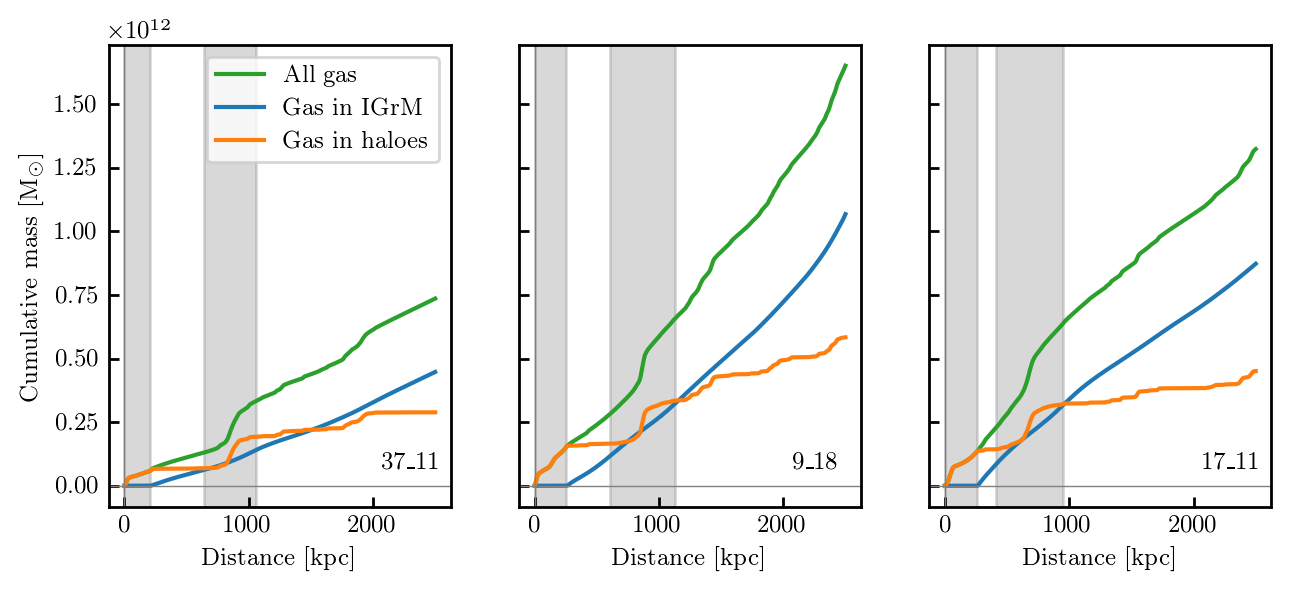}
    \caption{Cumulative mass of gas in the simulated LGs as a function of distance from Sun's position. The contributions of material belonging to the IGrM (i.e., excising material inside galaxy haloes) and within galaxy haloes only (i.e. including both the ISM and CGM of each galaxy but excluding the IGrM) are shown. The vertical grey shaded regions indicate the extent of MW and M31 virial radii.}
    \label{fig:radial_mass}
\end{figure*}

\noindent where $\vectorbold*{v}_{\rm g}$ is the peculiar velocity field of gas or galaxies in the simulations, $\vectorbold*{v}_{\rm MW}$ is the systemic velocity of the MW, $\vectorbold*{v}_{\odot}$ is the tangential velocity of the Sun around the MW centre and $\vectorbold*{v}_{\rm H}$ is the Hubble flow. For simplicity, we approximate the latter by $H_0\cdot\vectorbold*{r}$, where $\vectorbold*{r}$ is the radial vector of gas or galaxies from the observer's position. The GSR velocities are obtained by cancelling out the contribution of the Sun's tangential velocity in the LSR frame, namely,  

\begin{equation*}
\vectorbold*{v}_{\rm GSR} = \vectorbold*{v}_{\rm LSR} + \vectorbold*{v}_{\odot}.
\end{equation*}

Similarly, to obtain LGSR velocities, a correction for the radial motion of the MW with respect to the LG barycentre is applied to the GSR frame\footnote{We present this set of equations for $v_{\rm LSR}$, $v_{\rm GSR}$ and $v_{\rm LGSR}$ in vectorial form to emphasise the kinematics involved in each definition. Note that the scalar form in which these equations are usually presented in the observational context only involves the apex correction for Sun's velocity and the radial component of $\vectorbold*{v}_{\rm MW,LG}$ along the line of sight.}:  

\begin{equation*}
\vectorbold*{v}_{\rm LGSR} = \vectorbold*{v}_{\rm GSR} + \vectorbold*{v}_{\rm MW,LG} \cdot (\vectorbold*{x}_{\rm MW} - \vectorbold*{x}_{\rm LG}) \frac{\vectorbold*{x}_{\rm MW} - \vectorbold*{x}_{\rm LG}}{|\vectorbold*{x}_{\rm MW} - \vectorbold*{x}_{\rm LG}|^2},
\end{equation*}

%MW velocity wrt to LG barycentre:
%LG barycentre galactic coordinates ->
%IDL> l=147.*!PI/180. & b=-25.*!PI/180.
%IDL> print,l,b
%      2.56563    -0.436332
%IDL> print,-62.*cos(l)*cos(b) + 40.*sin(l)*cos(b)-35.*sin(b) = 81.6619 km/s

\noindent where $\vectorbold*{v}_{\rm MW,LG}$ is the physical velocity of the MW as seen from the LG barycentre and $\vectorbold*{x}_{\rm MW}$ and $\vectorbold*{x}_{\rm LG}$ are the MW and barycentre positions, respectively. The barycentre's position and velocity in each realization was computed considering all matter (gas, stars and DM) located within a radius of $1\,$Mpc from the midpoint between MW and M31; a distance that roughly defines the edge of the LG, as it will be shown in Section~\ref{sec:results}. %(see Fig.~\ref{fig:vrel_dist}). 

A visual impression of the gas around the MW, as viewed from the Sun's position, is presented in Fig.~\ref{fig:nh_density} for the three {\sc Hestia} high-resolution simulations. Both the H\,{\sc i} column densities {\it inside} and {\it outside} the virial radius of the MW galaxies are shown. The column density of neutral hydrogen atoms has been estimated using the ionization tables computed by \cite{Hani18}. The latter are built using the spectral synthesis code {\sc Cloudy} \citep[][]{Ferland17}, which computes equilibrium models to determine the ion fractions as a function of gas density, temperature and metallicity. The only ionising radiation field assumed in this calculation is the UV background of \cite{FaucherGiguere09}. Self shielding from the UV radiation in high-density regions is also included following \cite{Rahmati13}.

In the left-hand panels of Fig.~\ref{fig:nh_density}, we show material within the CGM of the MW candidate in each simulation, whereas material belonging to the LG and M31 can be seen in the right. A natural boundary for the CGM of galaxies in the simulations is that given by the virial radius of each halo, which we approximate here using $R_{200}$. The M31 partner of each simulation can be spotted in the upper-right corner of the sky maps for $37\_11$ and $17\_11$, and in the lower-right corner in the case of $9\_18$. The LG barycentre and anti-barycentre directions are also indicated, the former being, in general, close to M31's line of sight. 
In particular, in the three simulations, a gaseous filament reaching from the {\it general} barycentre direction to its antipole in the sky is seen in the right-hand panels. We note, however, that although general properties are shared by all the LGs, the unconstrained nature of the realizations at scales smaller than that of shell-crossing, i.e. well into the non-linear regime, makes each simulation unique.  

\subsection{Gas in the simulated LGs}

%textcolor{red}{\bf Acá la idea es describir el cálculo de las column densities brevemente basándose en el paper de Mitali y luego discutir la distribución de gas en en el CGM de los haloes y el IGM. Para eso conviene hacer la Fig.~\ref{fig:radial_mass} con los 3 plots cumulativos en forma horizontal ocupando todo el ancho de la página.}

Before moving into the kinematics of the LG gas it is interesting to assess the amount of gaseous material in the ISM and CGM of galaxies in comparison to that present in the IGrM. The cumulative gas mass as a function of distance to the observer can be seen in Fig.~\ref{fig:radial_mass}, where the joint contribution of all material {\it inside} and {\it outside} the virial radius of each galaxy is shown: i.e., we plot all gas belonging to the ISM and CGM of galaxies (orange solid lines), and of the IGrM only (blue solid lines), respectively. Within the simulated LGs (i.e., for distances $r\lesssim 1\,$Mpc), the amount of gas belonging to the ISM and CGM of the two main LG members is essentially of the same order of magnitude to that in the IGrM for all three high-resolution simulations and display similar radial profiles. As distance from the Sun's position increases, curves describing gas within haloes display a step-like behaviour, abruptly increasing after encountering a given galaxy. On the contrary, in the case of material in the IGrM, the curves show a smooth growth with distance, having roughly the same slope. Both simulations $9\_18$ and $17\_11$ are comparable in terms of radial profiles and gas content, amounting to a total of $M_{\rm g}(r\lesssim 1\,{\rm Mpc})\sim 6\times10^{11}\,$M$_{\odot}$, in contrast with simulation $37\_11$ that has a factor of $\sim2$ less gas within the LG. These differences are not surprising as they result from the expected cosmic variance of LG-sized volumes. However, irrespective of the particular realization analysed, these plots demonstrate that gas within the CGM and ISM of the MW and M31 galaxies is extremely relevant when accounting for the gas budget of LG-like systems, a fact that could have an impact on the interpretation of observations in our local environment, as it will be discussed in Section~\ref{sec:discussion}. At larger distances, material in the IGrM becomes the most relevant contribution to the total gas budget within a given sphere, as expected.  

\begin{figure*}
    \centering
    \includegraphics[width=2\columnwidth]{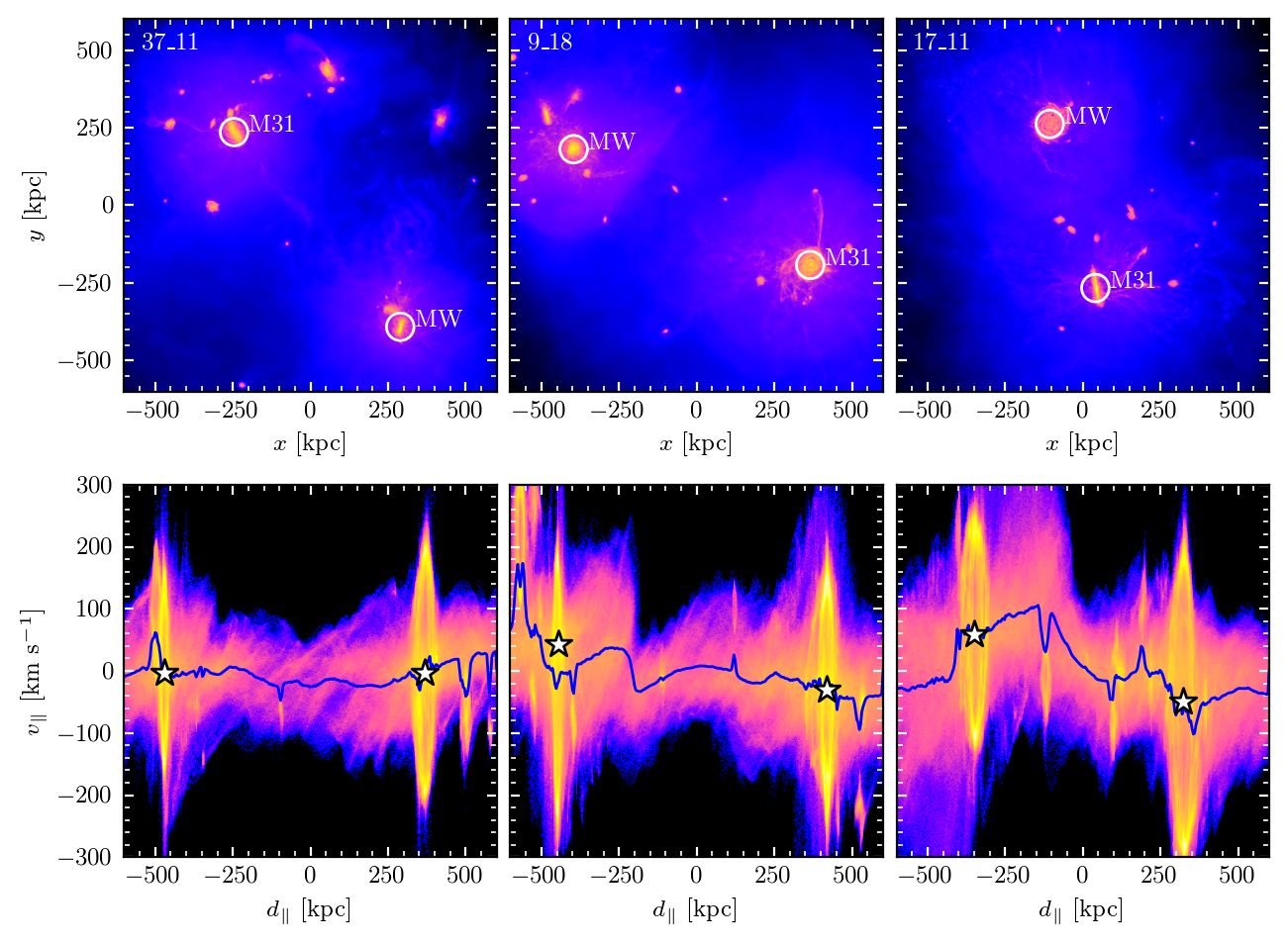}
    \caption{Projection of gas mass for each simulation (upper panels) and the distribution of gas cells in the $d_{\parallel}-v_{\parallel}$ plane (lower panels), for material within a distance of $300\,$kpc from the line joining the LG barycentre and M31's position. In the latter, the solid blue lines correspond to the median velocity values, whereas the white stars correspond to MW and M31 galaxy haloes considering the contribution of all matter within $R_{200}$.}
    \label{fig:dPar_vPar}
\end{figure*}

To further assess the general morphology and kinematics of the gaseous component in the simulated LGs, we define the quantities $d_{\parallel}$ and $v_{\parallel}$ in a similar manner to \cite{Sparre22}:

\begin{equation*}
    \centering
    d_{\parallel} \equiv \frac{(\vectorbold*{x}_{\rm{gas}} - \vectorbold*{x}_{\rm LG}) \cdot (\vectorbold*{x}_{\rm{M31}} - \vectorbold*{x}_{\rm LG})}{|\vectorbold*{x}_{\rm{M31}} - \vectorbold*{x}_{\rm LG}|} ,
\end{equation*}

\noindent where $\vectorbold*{x}_{\rm{gas}}$ is the position of the gas cell, $\vectorbold*{x}_{\rm{M31}}$ is the position of M31 and $\vectorbold*{x}_{\rm LG}$ the position of the LG barycentre. The corresponding parallel velocity is therefore defined as

\begin{equation*}
    \centering
    v_{\parallel} \equiv \frac{(\vectorbold*{v}_{\rm{gas}} - \vectorbold*{v}_{\rm LG}) \cdot (\vectorbold*{x}_{\rm{M31}} - \vectorbold*{x}_{\rm LG})}{|\vectorbold*{x}_{\rm{M31}} - \vectorbold*{x}_{\rm LG}|},
\end{equation*}

\noindent where $\vectorbold*{v}_{\rm{gas}}$ is the velocity of the gas cell and $\vectorbold*{v}_{\rm LG}$ is the physical velocity of the LG as seen from the MW. Therefore, these quantities measure the position and velocity of material in the LG with respect to the barycentre projected along the line joining the latter and M31. Note that these are both signed quantities: the MW projected position is negative while that of M31 is positive. Also, negative velocities run anti-parallel to the MW-M31 vector while the opposite is true for the positive velocities.

In Fig. \ref{fig:dPar_vPar}, we show projections of gas mass for each simulation (upper panels) and the distribution of gas cells in the $d_{\parallel}-v_{\parallel}$ plane (lower panels), for all gaseous material within a distance $d_{\perp}=300\,$kpc from the line joining the LG barycentre and M31's position. This means that the distributions seen in the lower panels correspond to gas cells inside a cylinder containing both the MW and M31 galaxies, up to distances that comprise the extension of their respective virial radii. In these plots, the Hubble flow is added from the position of the LG barycentre before projecting onto the $d_{\parallel}$ axis. 

The MW and M31 form two distinct regions in the $d_{\parallel}-v_{\parallel}$ plane, being the MW the biggest spot with $d_{\parallel} < 0$ and M31 the biggest one with $d_{\parallel} > 0$. For comparison, in each simulation, we also include white star symbols indicating both the MW and M31 galaxy haloes with $d_{\parallel}$ and $v_{\parallel}$ values obtained after averaging all material within $R_{200}$, i.e. considering not only gas, but also the contribution of the dark matter and stellar components. The observed difference in $v_{\parallel}$ between both galaxies provides a clear picture of the overall kinematics in each realization. In $17\_11$ (lower-right panel), the pronounced negative gradient in $v_{\parallel}$ indicates an overall infalling motion of both galaxies, which is less abrupt in $9\_18$ (lower-central panel), for which the infall velocity is smaller, and virtually inexistent in $37\_11$ (lower-left panel), where both galaxies are, in fact, temporarily receding (though close to turnaround). From these plots, one can infer that the simulation with kinematic properties more similar to the actual LG is $17\_11$, as it will be shown below.

\section{Results}
\label{sec:results}

In this section, we want to investigate the relation between the kinematics of gas and galaxies in the simulated LGs and their adjacent regions. It is well known that LG galaxies are part of a general flow towards the LG barycentre \citep[e.g.,][]{Peebles11,Whiting14}. Establishing the relation between gas and galaxy motions in the LG is, therefore, an important task, since the kinematics of high-velocity absorber samples in the CGM of the MW and its surroundings could potentially provide information on the global kinematics of the group (R17). %We start by analysing the resulting velocity pattern from the simulated LSRs, to later exclude the rotation of the galaxy from the analysis. 
In what follows, we study the velocity patterns seen by a set of observers for the three high-resolution {\sc Hestia} simulations, starting with LSR maps, to later move to the GSR and LGSR frames. For the gas component, we only consider the observed bulk velocity of material located along the respective line of sights without discriminating among different species. The contribution of different ions to the gas flows at $z=0$ will be extensively studied in a follow-up paper.  

\subsection{Simulated LSR maps}
\label{sec:LSR_maps}

\begin{figure}
	%\includegraphics[width=\columnwidth]{figures/vel_rad_rectangle_Hestia09-18_LSR_False.pdf}
	%\includegraphics[width=\columnwidth]{figures/vel_rad_rectangle_Hestia17-11_LSR_False.pdf}
	%\includegraphics[width=\columnwidth]{figures/vel_rad_rectangle_Hestia37-11_LSR_False.pdf}
	%\hspace*{-1cm}\includegraphics[width=1.3\columnwidth]{figures/vel_rad_rectangle_Hestia09-18_LSR_True.pdf}
	%\hspace*{-1cm}\includegraphics[width=1.3\columnwidth]{figures/vel_rad_rectangle_Hestia17-11_LSR_True.pdf}
	%\hspace*{-1cm}\includegraphics[width=1.3\columnwidth]{figures/vel_rad_rectangle_Hestia37-11_LSR_True.pdf}
	\vspace{0.5cm}
    \hspace{-0.6cm}
    \includegraphics[width=1\columnwidth]{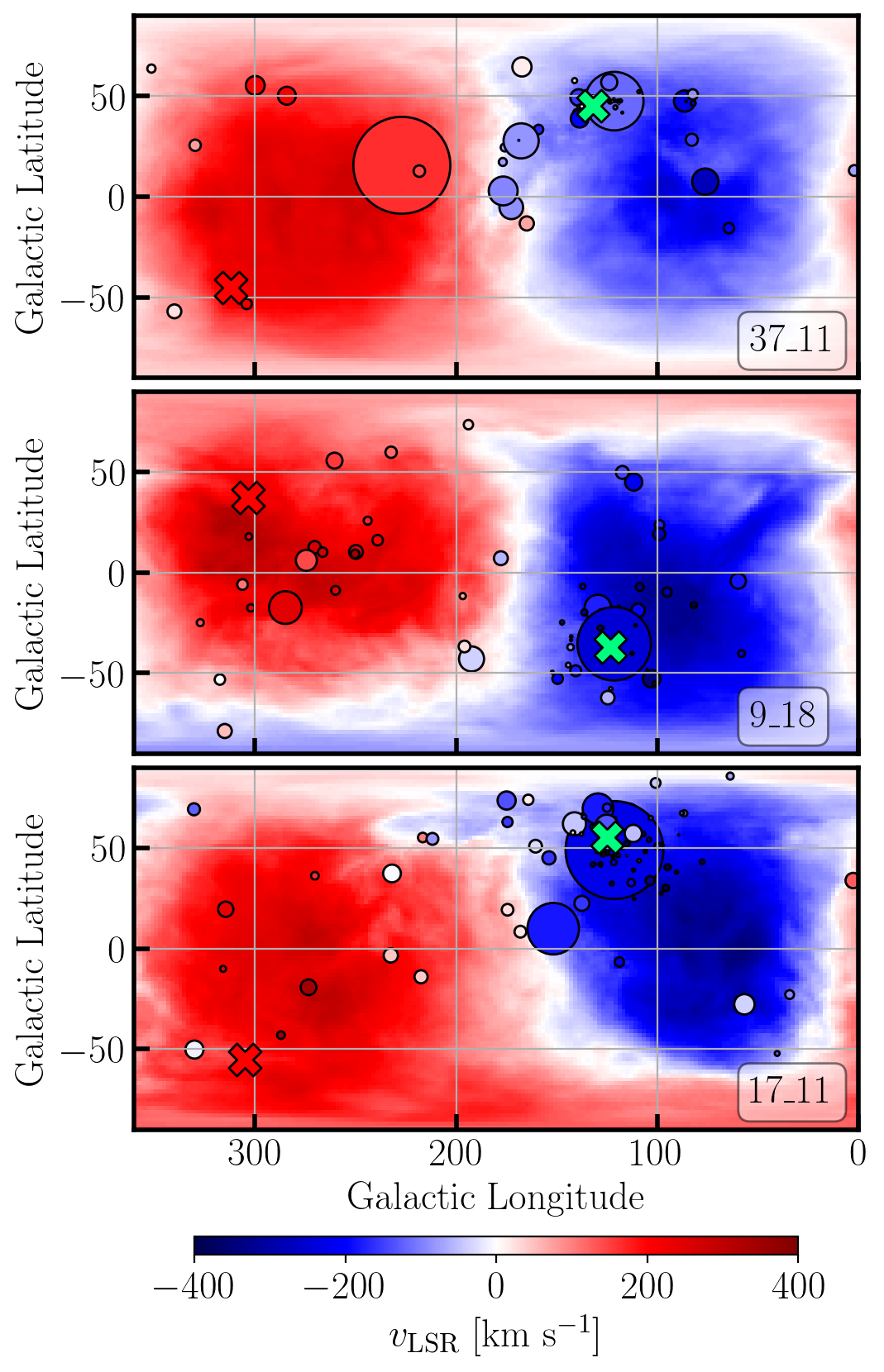}\vspace{-0.2cm}
    \caption{Sky distribution of galaxies with $L_V > 10^4\, L_{V_\odot}$ {\it outside} the virial radius of the MW analog for the simulated LGs. Solid circles are colour-coded by LSR velocity with radius indicating a size of 2$R_{200}$ for each halo (see also R17, their Fig. 10, for a similar
  plot based on observed LG galaxy parameters). The mean mass-weighted gas velocity at any given line of sight is also shown.  Crosses indicate the LG barycentre (green) and anti-barycentre (red) directions.}
    \label{fig:LSR_sky}
\end{figure}

\begin{figure}
    \hspace{-0.5cm}
    \vspace{-0.5cm}
    \includegraphics[width=1.05\columnwidth]{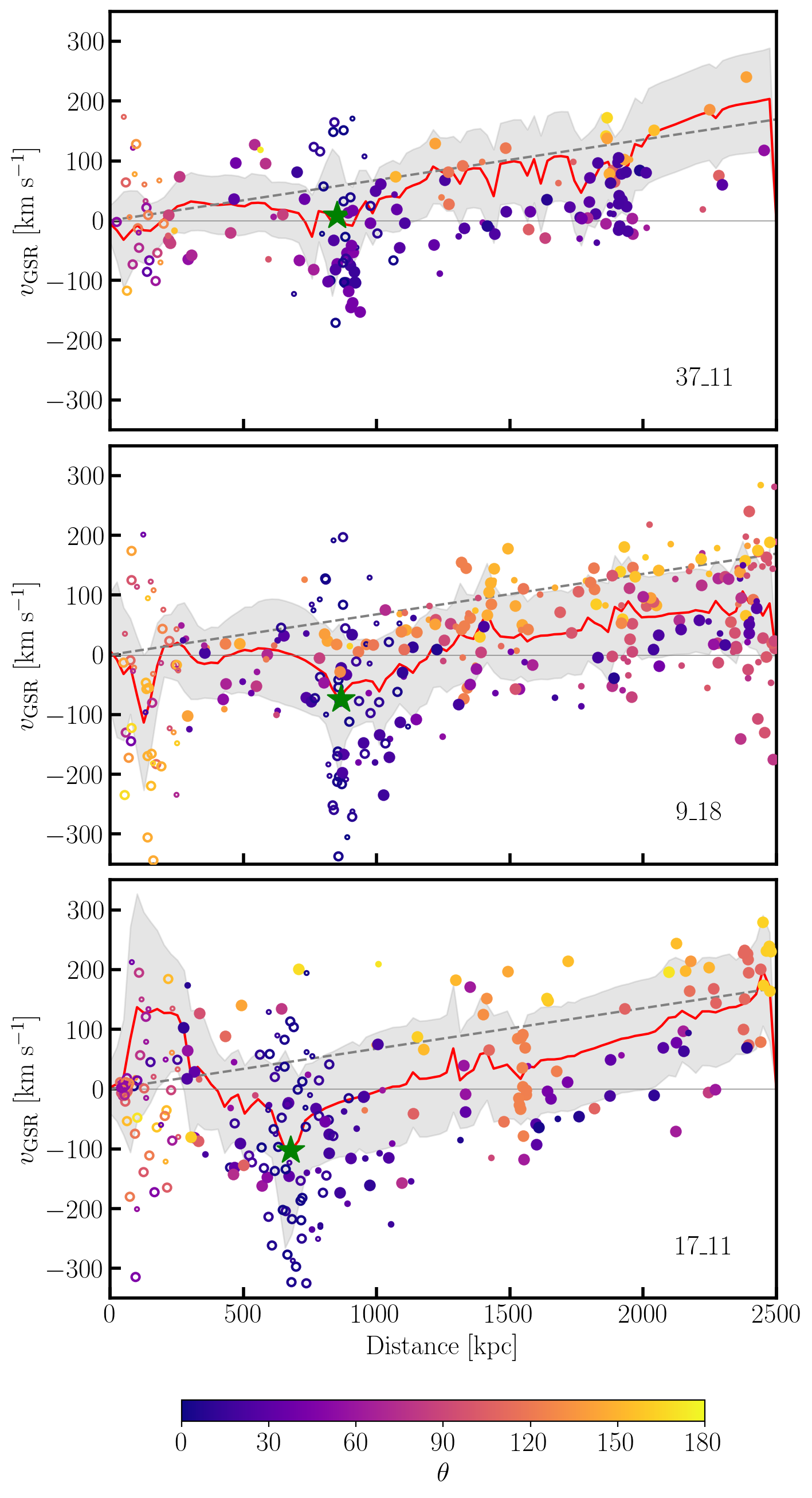}
    \caption{
    Mean mass-weighted radial velocity of all gas (red solid lines) and radial velocity of galaxies with $L_V > 10^4\,L_{V_{\odot}}$ (circles) in the GSR versus distance to the centre of the MW analog for the three high-resolution {\sc Hestia} runs. The grey shaded regions represent the $1\sigma$ standard deviation of the mean gas velocity. Galaxies are colour-coded according to the angle formed between their line of sight and M31's direction. Empty circles indicate satellites within the virial radius of MW (located at $r=0$) and M31 (shown as a green star). For comparison, the Hubble flow $v_{\rm H}=H_0 r$ is shown as a dashed line. 
    }
    \label{fig:vrel_dist}
\end{figure}

\begin{figure*}
    \centering
	\includegraphics[width=1.9\columnwidth]{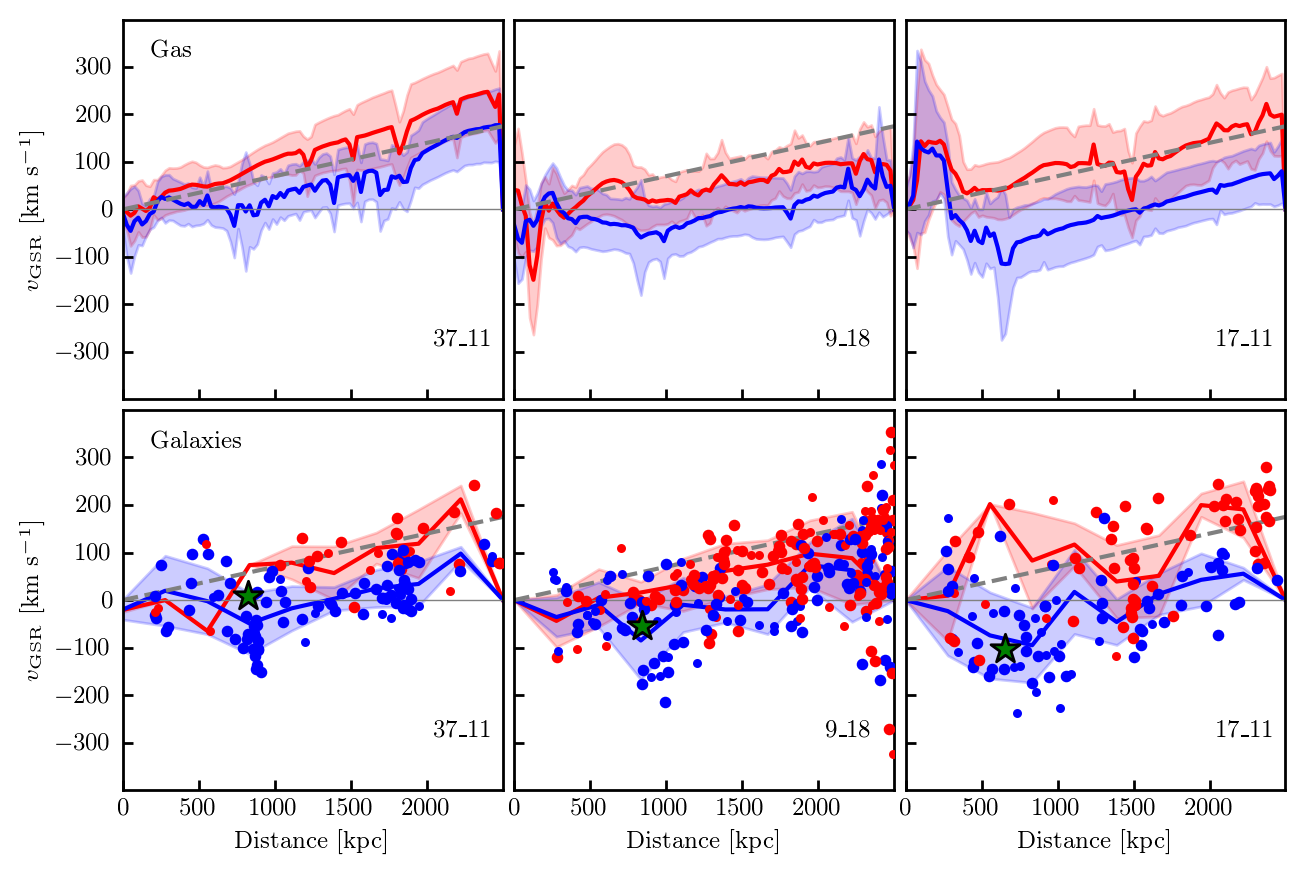}
    \caption{Idem as Fig.~\ref{fig:vrel_dist} but for gas and galaxies located {\it in front} ($\theta < 90^{\circ}$) and {\it behind} ($\theta > 90^{\circ}$) the MW in the GSR (blue and red, respectively). Galaxy satellites are excluded. For comparison, the Hubble flow $v_{\rm H}=H_0 r$ is shown as a dashed line.}
    \label{fig:vrel_back_front}
\end{figure*}

\iffalse
\begin{figure*}
	\includegraphics[width=2\columnwidth]{figures/lum_vs_mstar_Hestia17-11.png}
    \caption{Luminosidad en función de la masa estelar para 17-11
    }
    \label{fig:extra}
\end{figure*}
\fi

Fig.~\ref{fig:LSR_sky} shows the sky distribution of satellite galaxies viewed from the LSR (colour-coded by radial velocity) that are {\it outside} the MW's virial radius. To increase statistics, we plot all galaxies formed in the high-resolution simulated LGs that have luminosities an order of magnitude lower than the observational limit of $L_V\sim 10^5\,L_{\odot}$ given by \cite{Yniguez14} for LG satellites (symbols sizes are proportional to galaxy's $R_{200}$). This low-luminosity cutoff translates on an average galaxy stellar mass of $\sim 2\times10^4\,$M$_{\odot}$. To compute the LSR gas velocity, we perform a mass-weighted average of all gas cells along the line of sight. The distribution of gas radial velocity displays a perceptible velocity dipole pattern in all three LG simulations, resulting from the combination of the MW's galaxy rotation and its motion towards the LG barycentre. Generally speaking, most of the material at $l\lesssim 180^{\circ}$ is {\it approaching} the observer, whereas the opposite is true at higher longitudes. Moreover, most of the galaxies follow the same trend as the gas, although, as expected from the intrinsic random component of galaxy velocity vectors, there are some deviations. Noteworthy to mention is the large concentration of galaxies with mainly negative LSR velocities that can be seen towards the direction of the LG barycentre, some of which are M31's satellites. The sharp velocity contrast seen between positive and negative longitudes is mainly produced by the rotation of the galaxy which, as a result, affects measurements from the LSR. This effect is, however, more relevant at latitudes close to the galactic plane. At higher latitudes, the relative motion between MW and M31 also imprints a dipole pattern in the sky whose strength and overall sign depends on the absolute velocity between the galaxies. As mentioned in Section~\ref{sec:sim_LG}, the relative velocity of the simulated MW-M31 systems is $+9, -74, -102$ km$\,$s$^{-1}$ for the $37\_11$, $9\_18$ and $17\_11$ realizations, respectively. Therefore, the general flow towards the LG barycentre is expected to be stronger in the simulation $17\_11$, which has a MW-M31 radial velocity resembling the observed value of $-109\pm4.4$ km$\,$s$^{-1}$ \citep{vanderMarel12}.

\subsection{Excluding galaxy rotation}
\label{sec:GSR_maps}

In Fig.~\ref{fig:vrel_dist}, we show the radial velocity of gas and galaxies from the GSR %(i.e., excluding galaxy rotation) 
as a function of distance to the MW for our three LG realizations. This is done to quantify the relative motion of gas and galaxies with respect to the MW as a whole. For the gas, we compute the mass-weighted mean radial velocity of all gas cells within concentric spherical shells centred on the position of the Sun. Within the virial radius of each MW candidate, gas velocity shows different behaviours. This is expected, as gas in these locations may be part of inflows (outflows) towards (from) the inner regions of the galactic halo. In the case of $37\_11$ and $9\_18$ simulations, {\it net} inflows are observed. On the contrary, in simulation $17\_11$, outflows dominate at $r\lesssim R_{200}$. In particular, the MW galactic disc of realization $17\_11$ is the most massive of the sample and, at $z=0$, it happens to be producing strong galactic winds. For simulations $9\_18$ and $17\_11$, the gas between the MW and M31 galaxies is, on average, either approaching or at rest with respect to the MW, whereas, for simulation $37\_11$, is receding from the observer owing to the positive relative velocity between the two main LG galaxies. As expected by the chosen mass of the two main halos in our simulated LGs, the zero velocity sphere of both gas and galaxies lies at a distance of roughly $1\,$Mpc (particular values depending on the realization), in line with the LG measurements of \cite{Karachentsev06}. At larger distances, the Hubble flow dominates over peculiar motions and the GSR radial velocity of gas and galaxies gradually approaches the universal expansion velocity given by the dashed line (as mentioned above, we approximate the Hubble flow by adding $H_0\cdot\vectorbold*{r}$ to the peculiar velocities).  

In Fig.~\ref{fig:vrel_dist}, galaxy symbols are colour-coded according to the angle $\theta$ subtended between the two directions pointing towards the galaxy and M31 from the Sun's position ($0^{\circ}\leq\theta\leq180^{\circ}$). Essentially, small (large) angles correspond to line of sights pointing towards (opposite to) M31's sky region. Therefore, using this definition, we can characterize the location of all gas and galaxies in relation to the observer, with the right angle separating the regions lying {\it in front} ($\theta < 90^{\circ}$) and {\it behind} ($\theta > 90^{\circ}$) the MW. In the case of simulation $17\_11$, the position of galaxies outside MW's virial radius with respect to the barycentre direction correlates with their observed velocity: most of the systems located behind (in front) the MW tend to show positive (negative) radial velocities. This trend can be seen building up progressively, within the zero velocity sphere, as the relative {\it approaching} radial velocity between the MW and M31 galaxies in the simulations increases (see panels from top to bottom in Fig.~\ref{fig:vrel_dist}).  At larger distances, most gas and galaxies show receding velocities in all directions owing to the Hubble drift, as expected.

The line of sight dependence with radial velocity for gas and galaxies located at distances $\lesssim 1\,$Mpc can be better seen in Fig.~\ref{fig:vrel_back_front}, where we plot velocities as a function of distance in the same way as in Fig.~\ref{fig:vrel_dist}, but discriminating all material located ahead of and behind the observer. The gas kinematics is particularly different in the simulations at $r\lesssim300\,$kpc from MW's centre, because in this region it is mainly dominated by MW Galactic winds or gas accretion. At greater distances, just as before, the trend gradually shows up in the different LG realizations as a function of relative MW-M31 radial velocity. In the case of realization $37\_11$ almost all gas, irrespective of its location with respect to the observer, shows positive velocities outside the MW virial radius; a fact that is consistent with LG kinematics in this particular simulation. For the other two simulations there is a clear dependence of gas radial velocity with line of sight, clearly demonstrating that material (most noticeable the gas component) in the vicinity of the MW tends to lag behind showing positive velocities in the GSR. In the barycentre direction, the MW halo rams into gas that is at rest at the barycentre or that flows towards it from the opposite side producing negative velocities. A similar situation is seen for the galaxies flowing towards the LG barycentre as shown in the panels below, although the limited number and complex dynamics of available LG galaxies makes the trends not so straightforward to interpret.

\begin{figure*}
    \centering
	\includegraphics[width=1.5\columnwidth]{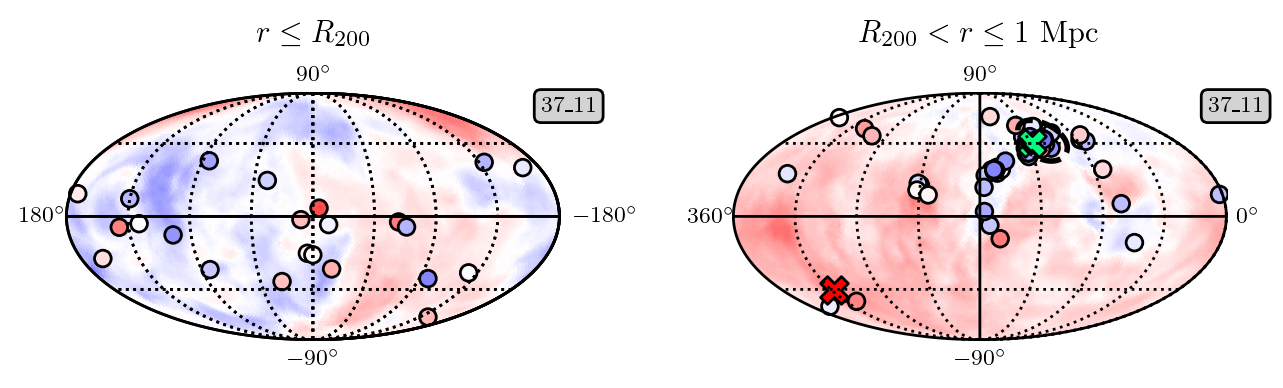}\vspace{0.3cm}
	\includegraphics[width=1.5\columnwidth]{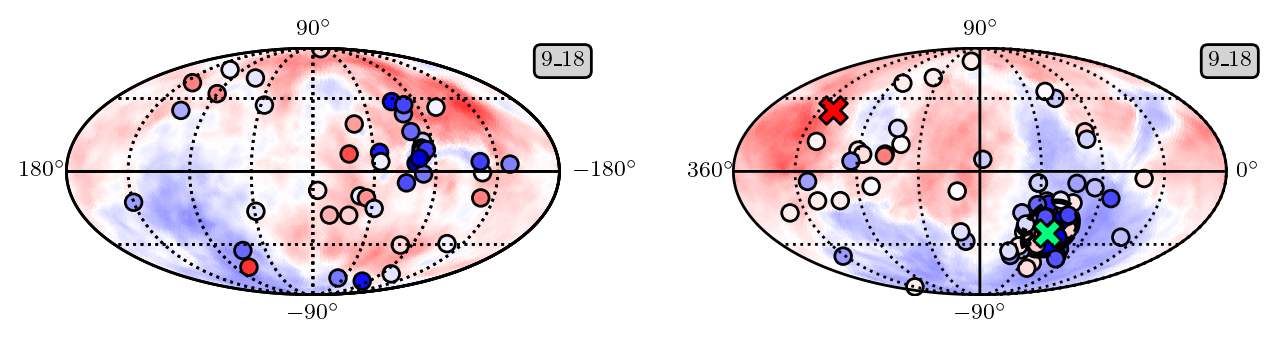}\vspace{0.3cm}
	\includegraphics[width=1.5\columnwidth]{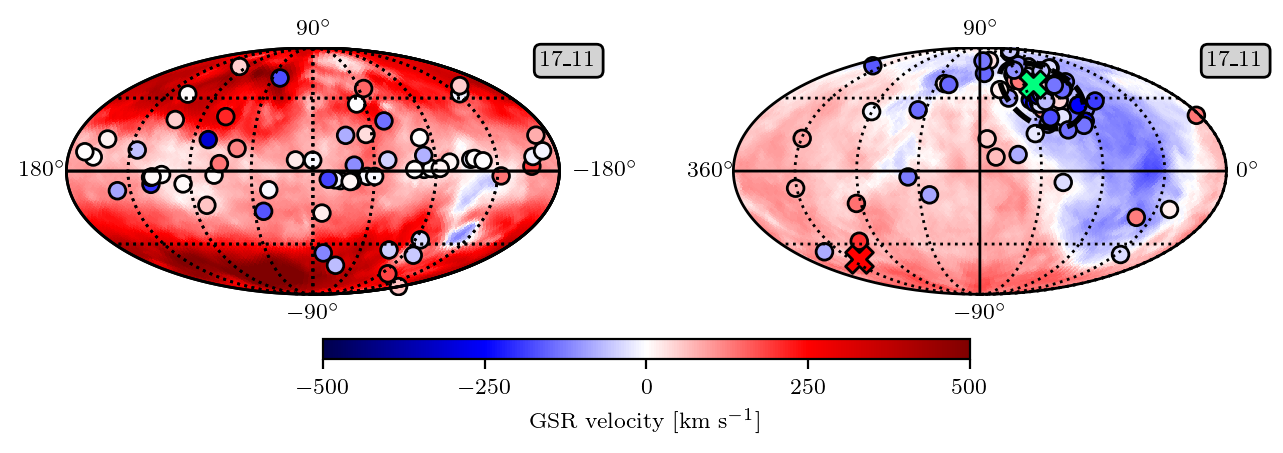}\vspace{0.3cm}
    \caption{Mollweide projection of mean mass-weighted gas and galaxies GSR velocities in the LG for the three {\sc Hestia} high-resolution runs at different distances from the Sun. Each panel also shows skymaps for all gas and galaxies at $r\leq R_{200}$ and $R_{200}<r\leq 1\,$Mpc ($R_{200}=205, 254, 255\,$kpc for the MW in $37\_11$, $09\_18$ and $17\_11$, respectively; see Table~\ref{table:1}). The green (red) cross indicates the barycentre (anti-barycentre) direction. Note the large concentration of galaxies towards the direction of M31 in all simulations. The black dotted line outlines the CGM of M31.}
    \label{fig:dist_bins}
\end{figure*}

The angular distribution of both gas and galaxies in the GSR from the inner galactic halo up to a distance of $1\,$Mpc is shown in Fig.~\ref{fig:dist_bins} for the three high-resolution {\sc Hestia} runs. The colour-coded radial velocities are computed in the same way as in Fig.~\ref{fig:LSR_sky}. The integrated angular distributions at $r \leq R_{200}$ and $R_{200}<r\leq1\,$Mpc are also shown. As noted in Fig.~\ref{fig:vrel_dist}, at $r\lesssim R_{200}$, the observed gas velocity is strongly linked to the central MW galaxy of each simulation; either by showing an average net accretion of material, or strong galactic winds. In the case of simulation $17\_11$, the gas is expelled out of the galactic plane mainly in the vertical direction, i.e. the heated gas tends to flow through the least-resistant medium, avoiding higher gas densities in the disc. This is clearly demonstrated by the low-radial velocity gas strip seen at $l=0$ in the $r\leq R_{200}$ sky projection (bottom left panel of Fig.~\ref{fig:dist_bins}). In the right panels, which correspond to $R_{200}<r\leq1\,$Mpc, a dipole pattern is shown to persist when removing Galactic rotation in simulations $9\_18$ and $17\_11$. Gas approaching the MW (blue) spans an area of the sky much bigger than the CGM of M31, which is outlined by the black dashed line. This suggests that the dipole velocity pattern as seen from the GSR in these two simulations (in which both galaxies are approaching) is related to the overall kinematics of the LG and cannot be attributed only to the CGM of M31. In this respect, we have checked that excluding the latter from the analysis does not affect our conclusions.

\subsection{Moving to the LGSR frame}
\label{sec:LGSR}

Owing to the lack of distance indicators in observations of high-velocity clouds it is often difficult to establish the origin of the absorbers, particularly for gas outside the CGM of the MW. Bracketing techniques using sources at known distances have been used in some particular cases but this method cannot be generally applied to the whole absorber population \citep{Ryans97,Wakker07,Wakker08,Thom06,Thom08, Lehner12, Lehner22}. An indirect method is to use the thermodynamic properties of the absorbing gas to infer its location within the LG. This type of analysis was done by \cite{Bouma19} focusing on directions close to the LG barycentre and its antipode, finding that
%, at least, 
a non-negligible fraction of the systems in the former are most likely located outside the MW virial radius, whereas the remaining systems (mainly seen towards the anti-barycentre region) could be either located within the MW halo or outside.

\begin{figure*}

	\includegraphics[width=2\columnwidth]{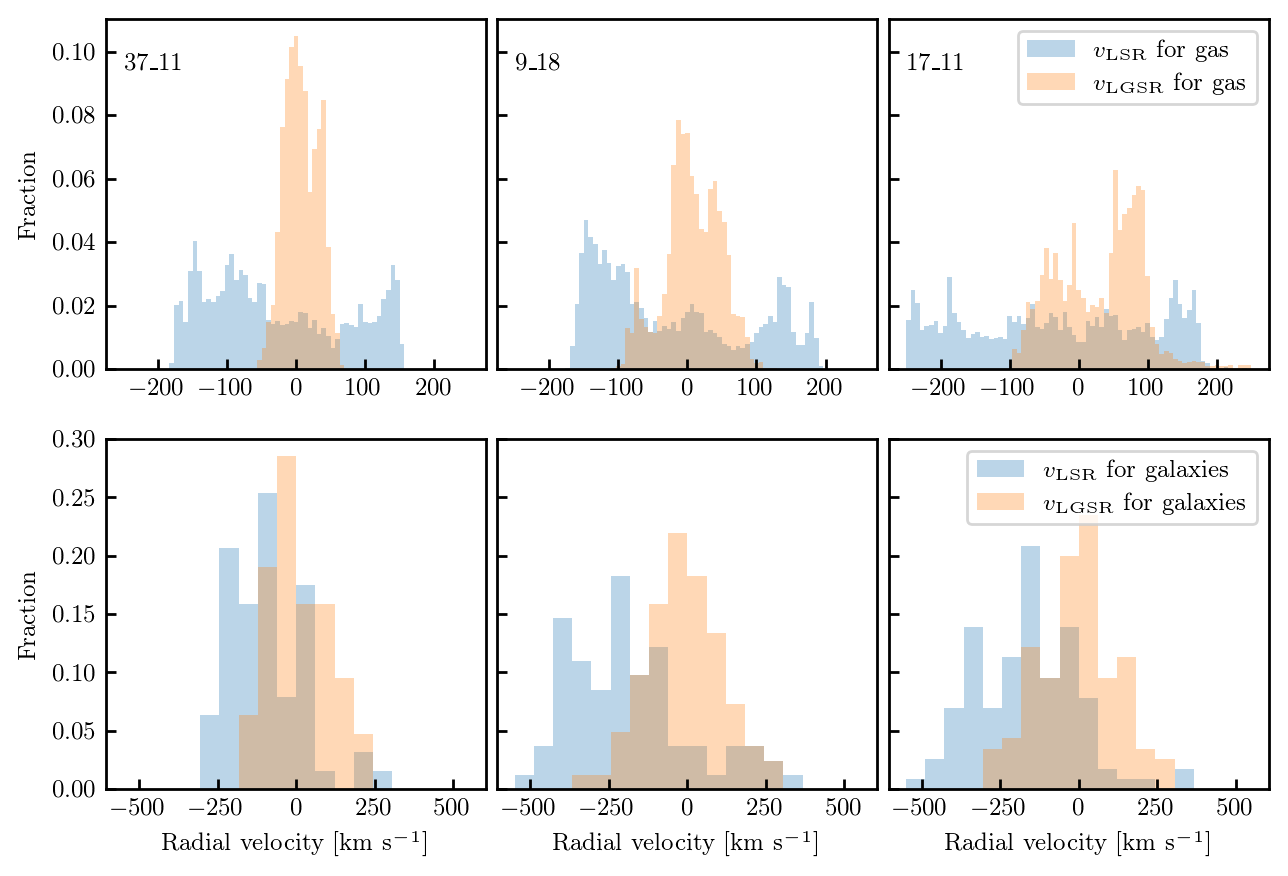}
    %\caption{LSR and LGSR velocity distributions for material inside \noindent where $\vectorbold*{v}_{\rm{gas}}$ is the velocity of the gas cell and $\vectorbold*{v}_{\rm LG}$ is the bulk velocity of the LG. Therefore, these quantities measure the position and velocity of material in the LG with respect to the barycentre projected along the line joining the latter and M31. Note that these are both signed quantities: the MW projected position is negative while that

%MW virial radius (upper panel) and outside (lower panel). We probe in directions distributed along the quadrants containing the barycentre and the antibarycentre.}
    \caption{LSR and LGSR velocity distributions for gas (top panels) and galaxies (bottom panels) outside the MW virial radius. For the gas, we probe 6080 directions evenly distributed along the quadrants containing the barycentre and anti-barycentre in each simulation.
    }
    \label{fig:LGSR_histo}
\end{figure*}

As stressed by \cite{Sembach03}, a necessary (but not sufficient) condition for the gas to be located outside the virial radius of the MW (and kinematically decoupled from it) is to observe a decrease in the spread of the radial velocity distributions as one moves from the LSR to the GSR and LGSR reference frames. To explore such an observable from the point of view of simulations we plot, in Fig.~\ref{fig:LGSR_histo}, the radial velocity distribution of gas and galaxies seen along the {\it general} barycentre and anti-barycentre directions in our three simulations. For comparison, we show distributions for both LSR and LGSR. The histograms are done selecting all line of sights found at two sky regions separated by the lines $l=180^{\circ}$ and $b=0^{\circ}$ (i.e., quadrants {\sc II} and {\sc IV} in simulation $9\_18$, and {\sc I} and {\sc III} in simulations $37\_11$ and $17\_11$, for the barycentre and anti-barycentre ``regions'', respectively) and are similar to those analysed by R17. %We have checked that varying the size of the selected regions does not dramatically affect our conclusions.     

As observed in the latter, from the point of view of the LSR, the distribution of gas shows a clear bimodality, stressing the fact that barycentre and anti-barycentre regions form a dipole in all simulations, as seen in Fig.~\ref{fig:LSR_sky}. After transforming the velocities to the LGSR, the standard deviation of the distributions noticeably decreases.
\iffalse
, spreading around zero in the case of simulations $9\_18$ and $17\_11$, but being mostly positive for $37\_11$, in line with the results of Fig.~\ref{fig:vrel_back_front}. This is expected because, as already mentioned above, the latter harbours a LG with MW and M31 galaxies receding from each other.
\fi
Interestingly, a dip test \citep{Hartigan85} suggests a clear bimodality for LGSR gas\footnote{Dip tests were run on $v_{\rm LGSR}$ gas distributions along the barycentre and anti-barycentre quadrants, yielding $p$-values smaller than $0.005$ in all simulations, which rejects unimodality at $\gtrsim 99.5\%$ certainty. From visual inspection, we suggest that these distributions are, in fact, primarily bimodal.}, strengthening the usual interpretation made by observers of LG gas flowing towards the barycentre for material seen towards these sky regions \citep[see e.g.][and references therein]{Bouma19}. For galaxies the situation is similar, although, for the three simulations, there is a large asymmetry in galaxy number counts between the two quadrants analysed. This is not surprising since directions close to the barycentre are mainly populated with M31 satellites, in contrast to the fewer LG member galaxies seen at its antipode that are outside the virial radius of the MW (see e.g. Table A.3 in R17). This is also seen in actual observations of LG galaxies and absorber populations. When transforming galaxy radial velocities from the LSR to the LGSR, there is a decrease of the standard deviation and skewness of the distributions, consistent to the results of R17. The relatively small amount of galaxies ($\sim100$) in the simulated LGs means we cannot make any reliable conclusion concerning bimodality in their velocity distribution\footnote{Galaxies along the barycentre and anti-barycentre quadrants on the LGSR frame yielded $p$-values of $0.97$, $0.77$ and $0.81$ for simulations $37\_11$, $9\_18$ and $17\_11$ respectively. These values suggest unimodality, but not with enough certainty to be taken as conclusive.}.

\section{Summary and Discussion}
\label{sec:discussion}

%In this work, we analysed the current motion of material in the LG as seen from a simulated observer at Sun's position to extend the previous analysis performed LG simulationrmed within the {\sc Clues} collaboration  

In this work, we have studied the kinematic properties of LG gas and galaxies in a suite of three high-resolution ($37\_11$, $9\_18$ and $17\_11$) simulations belonging to the {\sc Hestia} project \citep{Libeskind20} as seen from an observer located at Sun's position. Previous work using a LG simulation performed within the context of the {\sc Clues} collaboration has shown that the relative motion of the candidate MW and M31 galaxies determines a preferred direction within the LG that can affect both the kinematics and distribution of neighbouring material \citep[][]{Nuza14a,Richter17}. Our goal is to quantify the general flow of material towards the LG barycentre using the state-of-the-art {\sc Hestia} simulations to assess if this motion could provide a plausible explanation to the observed radial velocity dipole of high-velocity absorption systems towards the general barycentre direction and its antipode in the sky \citep[e.g.][]{Sembach03,Wakker03,Collins05,Richter17,Bouma19}. Several works have shown that this motion is also shared by LG galaxies outside the MW virial radius, a fact that is usually interpreted in the same way \citep[][]{Peebles89,Peebles01,Peebles11,Whiting14,Richter17}. To perform such a task, and compare results coming from gas and galaxies separately, we assessed the likelihood of finding a similar velocity dipole in our three LG simulations, which have a relative radial velocity of $+9, -74$ and $ -102\,$km\,s$^{-1}$ between the MW and M31 systems. This increase in relative velocity, from simulation $37\_11$ with its two main LG members receding from each other, to simulations $9\_18$ and $17\_11$ with approaching radial velocities, allows us to understand the impact of LG kinematics in the build up of the observed dipole at $z=0$.  

We have shown that the existence of a radial velocity dipole for both gas and galaxies outside the MW virial radius from the LSR is a natural outcome for all of our simulations (see Fig.~\ref{fig:LSR_sky}). This is expected as the rotation of the MW galaxy imprints a strong asymmetry in the sky maps, especially along the direction of the Sun's rotation vector, and close to the galactic plane. At high latitudes, galactic rotation plays a minor role and the resulting velocity maps are similar to those seen from the GSR, thus reflecting the {\it relative} motion between MW and M31. This effect, however, is more clearly seen in simulation $17\_11$ which has the largest relative radial velocity between them (see Fig.~\ref{fig:dist_bins}). Additionally, in this case, the relative velocity is very similar to the observed value reported by \cite{vanderMarel12} for the actual MW-M31 system. Interestingly, in Figs.~\ref{fig:vrel_dist} and~\ref{fig:vrel_back_front}, we have shown that, in simulation $17\_11$, a significant fraction of gas and galaxies outside of the MW virial radius located {\it in front} and {\it behind} the observer have {\it negative} and {\it positive} radial velocities from the GSR, respectively. These trends are stronger as one considers simulations with larger approaching MW-M31 relative velocities and, therefore, our findings are consistent with the interpretation given by R17 where the observer rams into LG gas at rest in the barycentre frame, with the MW moving faster than material that lags behind. 

When studying gas kinematics in the LGSR, the radial velocity distributions of gas outside the virial radius towards the barycentre/anti-barycentre quadrants in all simulations are in line with the results of R17, suggesting that the usual observational interpretation of gas and galaxies flowing towards the LG barycentre may be justified. Furthermore, the existence of gaseous material seen at these preferred line of sights that is part of the LG has been verified in some of the absorbers studied by \cite{Bouma19}, especially towards the barycentre region. 

There is, however, an important aspect to remark. \cite{Bouma19} found that many of the absorbers in their sample may also have originated in gas sitting at the edge or within the MW halo \citep[see also e.g.][]{Lehner12,Fox14,Richter17,Fox19,Marasco22,Lehner22}. If absorbing material displaying the observed dipole is located within the CGM of the MW, the actual scenario could be more complex than the simple kinematic interpretation discussed above, and it might depend on the peculiarities of gas flowing within the MW halo (see e.g. the different velocity flows seen at $r\leq R_{200}$ between simulations in Fig.~\ref{fig:dist_bins}). As discussed in R17, the Magellanic Stream and its large kinematics contribute to the observed dipole pattern of UV absorbers across the sky. While none of the high-resolution simulations presented here contain a feature that could be regarded as a Magellanic Stream analogue, Fig.~\ref{fig:radial_mass} indicates that galaxy haloes significantly contribute to the amount of absorbing material in the simulated LGs making the interpretation even more difficult. Therefore, it would not be surprising to expect that many of the absorbers actually correspond to gas located within the CGM of our Galaxy. In a follow-up paper we will study these aspects in detail also including the contribution of different ions to the total budget of LG gas.

\section*{acknowledgments}

SEN and CS are members of the Carrera del Investigador Cient\'{\i}fico of CONICET. They acknowledge funding from Agencia Nacional de Promoci\'on Cient\'{\i}fica y Tecnol\'ogica (PICT-201-0667).
RG acknowledges financial support from the Spanish Ministry of Science and Innovation (MICINN) through the Spanish State Research Agency, under the Severo Ochoa Program 2020-2023 (CEX2019-000920-S).
ET acknowledges support by ETAg grant PRG1006 and by EU through the ERDF CoE grant TK133.
JS acknowledges support from the French Agence Nationale de la Recherche for the LOCALIZATION project under grant agreements ANR-21-CE31-0019. 
NIL acknowledges support from the DFG research Project ``The Cosmic Web and its impact on galaxy formation and alignment'' (DFG-LI 2015/5-1).

%%%%%%%%%%%%%%%%%%%%%%%%%%%%%%%%%%%%%%%%%%%%%%%%%%
\section*{Data Availability}

The scripts and plots for this article will be shared on reasonable request to the corresponding author. The {\sc Arepo} code is publicly
available \citep{Weinberger20}.

\bibliographystyle{mnras}
\bibliography{biblio} % if your bibtex file is called example.bib

\label{lastpage}
\end{document}